\documentclass[twocolumn]{aastex6}
\usepackage{amsmath}
\usepackage[normalem]{ulem}

\begin{document}
\shorttitle{Crab Pulsar Scintillation}
\shortauthors{Main, R.}
\newcommand{\new}[1]{\textcolor{purple}{#1}}

\title{Resolving the emission regions of the Crab pulsar's giant pulses}
\author{Robert Main$^{1,2,3}$, Rebecca Lin$^{2}$,  Marten H. van Kerkwijk$^{2}$, Ue-Li Pen$^{3,5,4,6}$, Alexei G. Rudnitskii$^{7}$, Mikhail V. Popov$^{7}$, Vladimir A. Soglasnov $^{7}$, Maxim Lyutikov$^{8}$}

\affil{
$^{1}$Max-Planck-Institut f\"{u}r Radioastronomie, Auf dem H\"{u}gel 69, 53121, Bonn, Germany \\
$^2$Department of Astronomy and Astrophysics, University of Toronto, 50 St. George Street, Toronto, ON M5S 3H4, Canada \\
$^3$Canadian Institute for Theoretical Astrophysics, University of Toronto, 60 St. George Street, Toronto, ON M5S 3H8, Canada \\
$^4$Dunlap Institute for Astronomy and Astrophysics, University of Toronto, 50 St George Street, Toronto, ON M5S 3H4, Canada \\
$^5$Canadian Institute for Advanced Research, 180 Dundas St West, Toronto, ON M5G 1Z8, Canada \\
$^6$Perimeter Institute for Theoretical Physics, 31 Caroline Street North, Waterloo, ON N2L 2Y5, Canada \\
$^7$Astro Space Center, Lebedev Physical Institute, Russian Academy of Sciences, Profsoyuznaya ul. 84/32, Moscow, 117997 Russia \\
$^8$Department of Physics, Purdue University, 525 Northwestern Avenue, West Lafayette, IN 47907-2036, USA
}

\begin{abstract}

The Crab pulsar has striking radio emission properties, with the two dominant pulse components -- the main pulse and the interpulse -- consisting entirely of giant pulses.  The emission is scattered in both the Crab nebula and the interstellar medium, causing multi-path propagation and thus scintillation.  We study the scintillation of the Crab's giant pulses using phased Westerbork Synthesis Radio Telescope data at 1668\,MHz.  We find that giant pulse spectra correlate at only $\sim 2 \%$, much lower than the $1/3$ correlation expected from a randomized signal imparted with the same impulse response function.
In addition, we find that the main pulse and the interpulse appear to scintillate differently; the 2D cross-correlation of scintillation between the interpulse and main pulse has a lower amplitude, and is wider in time and frequency delay than the 2D autocorrelation of main pulses.
These lines of evidence suggest that the giant pulse emission regions are extended, and that the main pulse and interpulse arise in physically distinct regions which are resolved by the scattering screen.
Assuming the scattering takes place in the nebular filaments, the emission regions are of order a light cylinder radius, as projected on the sky.  With further VLBI and multi-frequency data, it may be possible to measure the distance to the scattering screens, the size of giant pulse emission regions, and the physical separation between the pulse components.
\end{abstract}

\section{The Unusual Properties of the Crab Pulsar}

The Crab pulsar is one of the most unusual radio pulsars, and has been the subject of much observational and theoretical research (for a review, see \citealt{eilek+16}). The two dominant components to its radio pulse profile, the main pulse and the low-frequency interpulse (simply referred to as the interpulse for the remainder of this paper), appear to be comprised entirely of randomly occurring giant pulses -- extremely short and bright pulses of radio emission showing structure down to ns timescales and reaching intensities over a MJy \citep{hankins+07}.  Only the fainter components of the pulse profile -- such as the precursor (to the main pulse) -- are similar to what is seen for regular radio pulsars.

The main pulse and interpulse are aligned within 2 ms with emission compenents at higher energy, from optical to $\gamma$-ray \citep{moffett+96,abdo+10}, and giant pulses are associated with enhanced optical \citep{shearer+03,strader+13} and X-ray \citep{enoto+21} emission. Since pair production strongly absorbs $\gamma$-ray photons inside the magnetosphere, this suggests all these components arise far from the neutron-star surface, with possible emission regions being the various magnetospheric ``gaps'' \citep{romani+95, muslimov+04, qiao+04, istomin04}, induced Compton scattering in the upper magnetosphere \citep{petrova04, petrova09}, or regions outside the light cylinder \citep{philippov+19}. In these regions, the giant pulses are thought to arise stochastically, likely triggered by plasma instabilities and/or reconnection \citep{eilek+16,philippov+19}, from parts smaller, of order $\Gamma c \tau_{\rm pulse}\simeq0.1\ldots1\,$km (with $\tau_{\rm pulse}\sim10\,$ns the timescale of a pulse, and $\Gamma\sim100$ an estimate of the relativistic motion), than the overall size of the emission region, of order $cP\delta\phi_{\rm pulse}\sim100\,$km (with $\delta\phi_{\rm pulse}\sim0.01$ the width of the pulse phase window in which giant pulses occur).

While similar in their overall properties, the main pulse and interpulse have differences in detail.  In particular, the interpulse has a large scatter in its dispersion measure compared to the main pulse, possibly suggesting that it is observed through a larger fraction of the magnetosphere \citep{eilek+16}.  In addition, it appears shifted in phase and shows ``banding'' in its power spectra above 4\,GHz (the so-called ``high-frequency interpulse''), with the spacing proportional to frequency \citep{hankins+07,hankins+16}.

The Crab pulsar, like many pulsars, exhibits scintillation from multi-path propagation of its radio emission.
The scattering appears to include both a relatively steady component, arising in the interstellar medium, and a highly variable one, originating  in the the Crab nebula itself, with the former responsible for the angular and the latter for (most of) the temporal broadening (\citealt{rankin+73, vandenberg+76b,popov+17,rudnistkii+17, mckee+18}).

This scintillation offers the prospect of ``interstellar interferometry'', where the high spatial resolution arising from multiple imaging is used to resolve the pulsar magnetosphere.  This has been applied to some pulsars, with separations between emission regions inferred from time offsets (or phase gradients) between the scintillation patterns seen in different pulse components.  For some of these pulsars, the inferred separations were substantially larger than the neutron star radius: $\sim\!10^{3}$\,km for PSR B1237+25 \citep{wolszczan+87}, $\gtrsim\!100\,$km for PSR B1133+16 \citep{gupta+99}, and of order the light cylinder radius (several $10^{4}$\,km) in a further four pulsars \citep{smirnova+96}.  In contrast, for PSR B0834+06, \citet{pen+14} find only a very small positional shift, constraining the separation between emission regions to $\sim\!20\,$km, comparable to the neutron star radius.

In the above studies, the scintillation pattern offsets are small compared to the scintillation scale, i.e., the scintillation screen does not resolve the pulsar magnetosphere, but changes in position can be measured with high signal-to-noise data.
For the Crab pulsar, however, the proximity of the nebular scattering screen to the pulsar implies that, as seen from the pulsar, the screen extends a much larger angle than would be the case if it were far away (for a given scattering time). Therefore, the scintillation pattern is sensitive to small spatial scales, of order $\sim\!2000\,$km at our observing frequency (see Sect.~\ref{sect:resolution}), comparable to the light-cylinder radius $r_{\rm LC}\equiv cP/2\pi\simeq1600\,$km.

The high spatial resolving power also implies that, for a given relative velocity between the pulsar and the screen, the scintillation timescale is short. Indeed, from the scintillation properties of giant pulses, \citet{cordes+04} infer a de-correlation time of $\sim\!25\,$s at $1.4\,$GHz. Unfortunately, their sample, while very large, had insufficient interpulse-main pulse pairs to look for differences between the two components (Cordes, 2017, pers.\ comm.).  From an even larger sample,
\citet{karuppusamy+10} identified pairs of pulses that either occurred in the same main pulse phase window, or with one in the main pulse and one in the interpulse window.  They did not find major differences between the sets.  Like for the close pairs in \citet{cordes+04}, they found correlation coefficients consistent with the $1/3$ expected for pulses that differ in their intrinsic time and frequency structure, but which have additional frequency structure imposed by scintillation.

In this paper, we compare the scintillation structure of the main pulse and the interpulse in more detail.  We find that in our sample the frequency spectra of close pulses are much more weakly correlated than was seen previously, suggesting that during our observations
the scintillation pattern is sensitive to smaller spatial scales at the source than the separation between bursts (in effect, the scattering screen resolved the emission region)
The scintillation patterns of the main pulse and interpulse also appear to differ, which, if taken at face value, suggests their emission locations are offset in projection by of order a light cylinder radius.

\section{Observations and Data Reduction}
\label{sect:data}

We analyse 6 hours of data form the phased Westerbork Synthesis Radio Telescope (WSRT), and 2.5 hours of simultaneous data from the 305$-$m  William  E.  Gordon  Telescope  at the Arecibo  observatory (AR), that were taken as part of a RadioAstron observing run on 2015 January 10--11 \citep{popov+17}. The data cover the frequency range of 1652--1684\,MHz, and consist of both circular polarizations in two contiguous 16\,MHz channels, recorded using standard 2-bit Mark~5B format (WSRT), and VDIF (AR).  The use of a telescope with high spatial resolution is particularly beneficial in studies of the Crab pulsar, as it helps to resolve out the Crab nebula, effectively reducing the system temperature from $830\,$Jy (for the integrated flux at 1.7\,GHz) to $165\,$Jy and $275\,$Jy, for WSRT and AR, respectively \citep{popov+17}.

\begin{figure*}
\center{}
\includegraphics[trim=0cm 0cm 0cm 0cm, clip=true, width=0.9\textwidth]{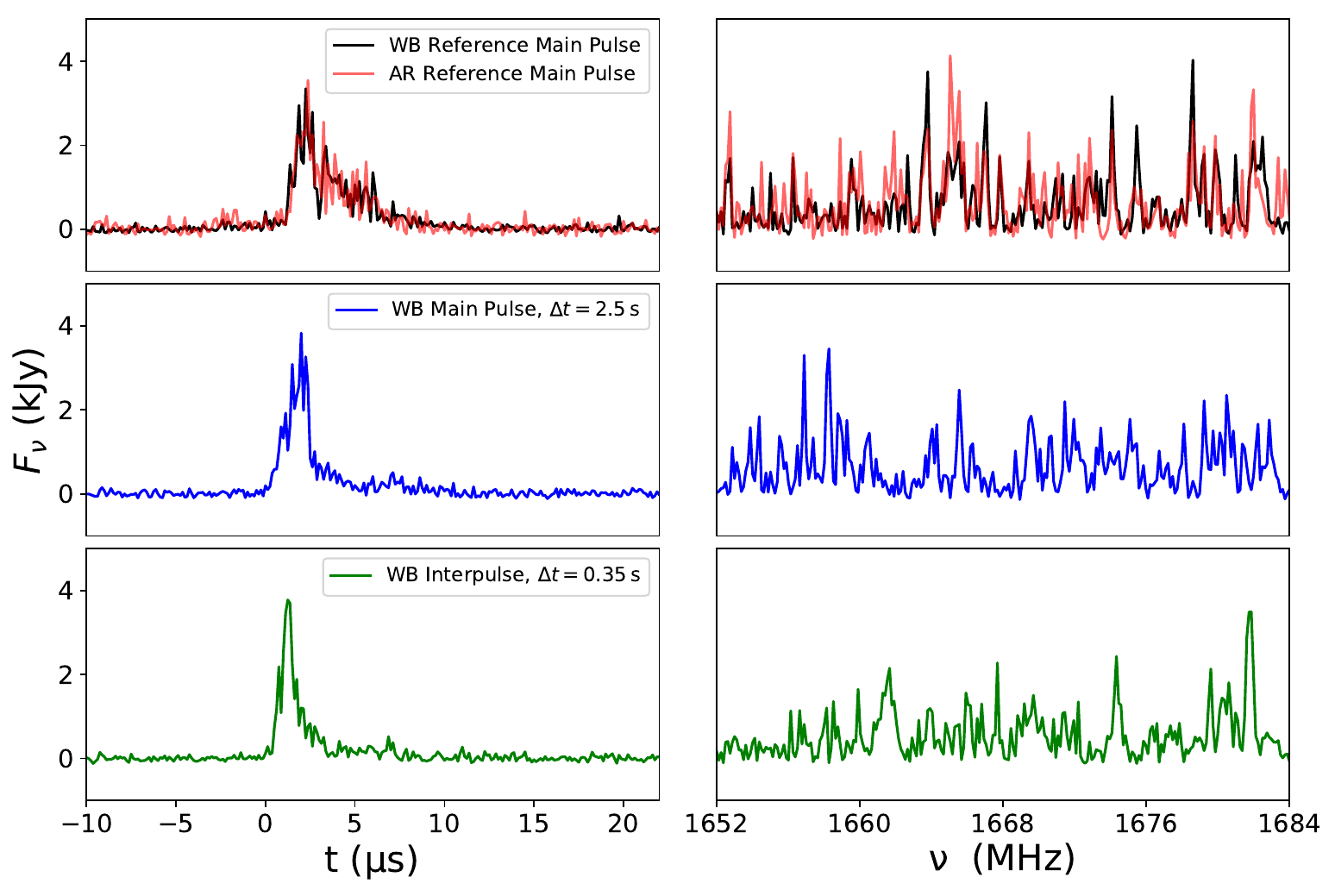}
\caption[Giant pulses of the Crab pulsar]{
{\em Top:\/} Pulse profiles ({\em left\/}) and spectra ({\em right\/}) of a main pulse at both WSRT and AR.  A nearby main pulse ({\em middle\/}) and interpulse ({\em bottom\/}) are shown for comparison, both separated by less than the scintillation timescale measured in section \ref{sec:corr}. The profiles are shown in 125\,ns bins, and total flux is calculated using the $T_{\rm{sys}}$ values in section \ref{sect:data}. The spectra contain the emission from 0--8\,$\mu$s, in 125\,kHz channels. The same pulse at WSRT and AR show clearly similar temporal and spectral structure, while the spectra of pulses within a scintillation time are drastically different, beyond what can be explained through the variable intrinsic structure in the pulse (discussed in detail in section \ref{sec:low}).
}
\label{fig:pulses}
\end{figure*}

To search for giant pulses, we coherently dedispersed\footnote{Using a dispersion measure of $56.7716{\rm\,pc\,cm^{-3}}$ appropriate for our date (taken from \url{http://www.jb.man.ac.uk/~pulsar/crab.html}; \citealt{lyne+93}). We read in overlapping blocks of data, removed edges corrupted by de-dispersion, such that the de-dispersed data was contiguous in time.}
the data from the two channels to a common reference frequency.  Each 1\,s segment of data was bandpass calibrated by channelizing the timestream into 8192 frequency channels per subband, normalizing by the square root of the time-averaged power spectrum. This correction works sufficiently well everywhere but the band edges. RFI spikes above $5\sigma$ are removed using a 128-channel median filter and time-variability is normalized by the square root of the frequency-averaged power spectrum. The signal was converted to complex by removing negative frequency components of the analytic representation signal (via a Hilbert transform), and shifting the signal down in frequency by half the signal bandwidth.
After forming power spectra, we replace the outer 1\,MHz (1652--1653\,MHz, 1683--1684\,MHz) and central 0.5\,MHz (1667.75--1668.25\,MHz) with the mean intensity in that time segment.

We search for giant pulses in a rolling boxcar window of $8\,\mu$s in steps of 62.5\,ns (one sample in the complex timestream), summing the power from both channels and both polarizations.  We flagged peaks above $8\sigma$ in the WSRT data, corresponding to $\sim\!60\,$Jy, as giant pulses, finding 15232 events, i.e., a rate of $\sim\!0.7{\rm\,s^{-1}}$. This detection threshold was chosen to assure there were no spurious detections. We find 4633 pulses above $8\sigma$ in the overlapping 2.5\,hours of AR data, and all pulses have a detectable, higher S/N counterpart in WSRT.  We show a sample main pulse detected at both telescopes, as well as main pulse and interpulse nearby in time in Fig.~\ref{fig:pulses}.

One possible concern is the effects of saturation from 2--bit recording, as described in \citealt{jenet98}.  The dispersion sweep in our frequency range of 1652--1684\,MHz is 3.26\,ms, $\sim$100000 samples. Thus, even the strongest giant pulse, having peak flux density of $\sim$100\,kJy and duration of 3\,$\mu$s will be reduced in intensity by roughly a thousand times, increasing the system temperature dS/S by only 35\% and 60\% for AR and WSRT respectively, with the recording systems far from saturation. The majority of our pulses are much fainter, near our detection threshold of $\sim\!60\,$Jy, where saturation effects will be negligible.

\section{Scintillation Properties}

With the phased WSRT array, our pulse detection rate is sufficiently high that it becomes possible to compute a traditional dynamic spectrum by summing intensities as a function of time.  We do this first below, as it gives an immediate qualitative view of the scintillation.  A more natural choice for pulses which occur randomly in time, however, is to parametrize variations as a function of $\Delta t$, the time separation between pulses \citep{cordes+04,popov+17}. Hence, we continue by constructing correlation functions of the spectra, as functions of both time and frequency offset.

\subsection{The Dynamic Spectrum of the Main Pulse}

\begin{figure*}
\begin{center}
\includegraphics[trim=2cm 4cm 4cm 3cm, clip=true, width=1.0\textwidth]{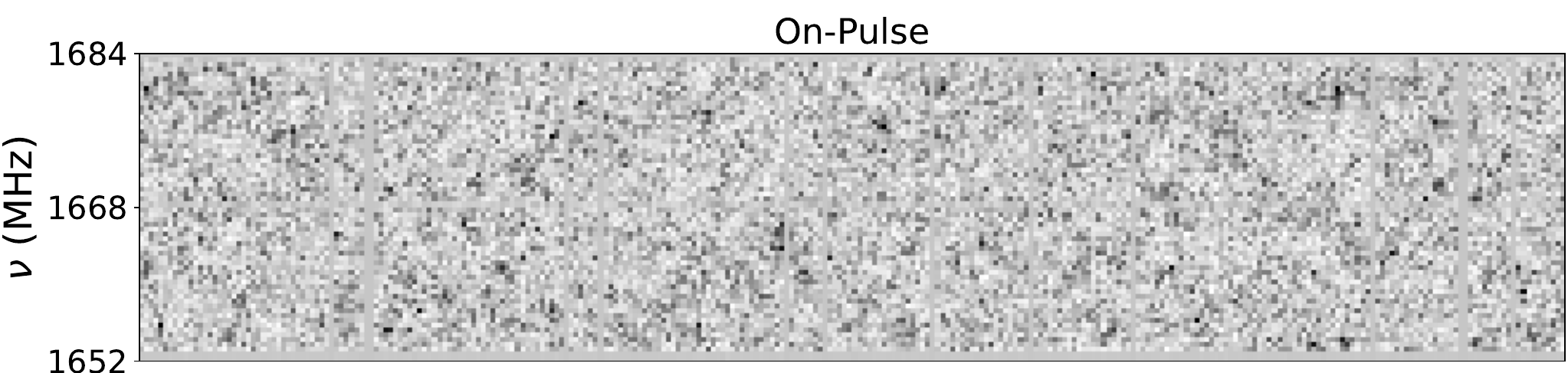}
\includegraphics[trim=2cm 2cm 4cm 3cm, clip=true, width=1.0\textwidth]{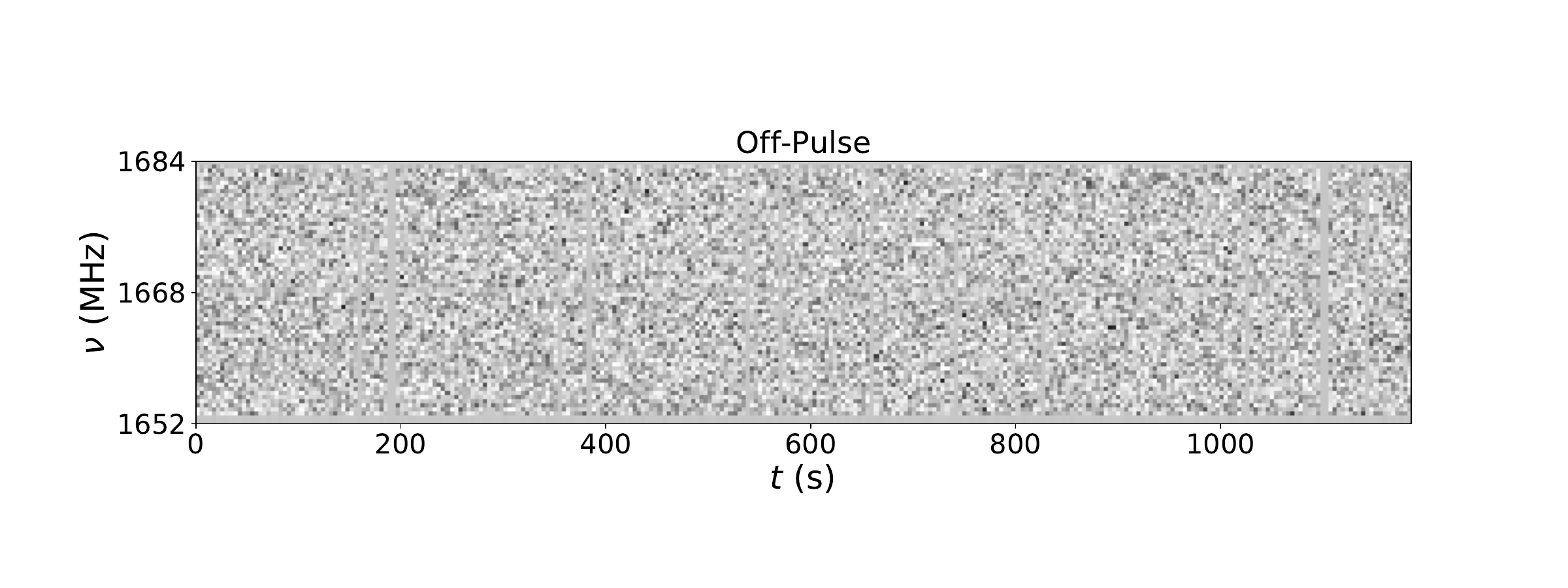}
\caption[Dynamic spectrum of the main pulse.]{{\em Top:\/} Part of the dynamic spectrum inferred from the main pulse by summing individual giant pulse spectra at 250\,kHz resolution in 4\,s bins.  The total flux in each time bin was normalized to remove the effects of variable pulse brightness. The random occurrence of giant pulses and their variable flux means that the noise properties of the time bins are heterogeneous, and that some bins have no flux. {\em Bottom:\/} Dynamic spectrum of an off-pulse region before each giant pulse, processed identically to the top panel. Artefacts from the bandpass are apparent, both at 1668\,MHz, and at the band edges, which are masked in further analysis.  The off-pulse region shows heterogeneous noise due to the variable number of ``pulses'' in each time bin, reflecting the uneven sampling of giant pulses. The colorbars are saturated to $\mu_{- 1\sigma}^{+2 \sigma}$, where $\mu$ and $\sigma$ are the mean and standard deviation of the on-pulse dynamic spectrum.
\label{figure:dynspec}}
\end{center}
\end{figure*}

We construct the dynamic spectrum $I(t, \nu)$ by simply summing giant pulse spectra in each time bin. Since we wish to observe the scintillation pattern rather than the vast intrinsic intensity variations between giant pulses, we normalize each time bin by the total flux within that bin. While there will still be structure in the dynamic spectrum owing to the intrinsic time structure of the giant pulses \citep{cordes+04}, any features in frequency which correlate in time should only be associated with scintillation.  We show a 20 minute segment of the dynamic spectrum in Figure~\ref{figure:dynspec}.  While noisy, the dynamic spectrum shows scintillation features. They are resolved by our time and frequency bin sizes of 4\,s and 250\,kHz, respectively, but only by a few bins, suggesting that the scintillation timescale and bandwidth are larger than our bin sizes by a factor of a few (consistent with $\nu_{\rm decorr}=1.10\pm0.02\,$MHz, $t_{\rm scint}=9.2\pm0.13\,$s measured below).

\subsection{Correlation Functions}
\label{sec:corr}
To infer the scintillation bandwidth and timescale, one usually uses the auto-correlation of the dynamic spectrum, but for pulses randomly spaced in time, it is easier to correlate spectra of pulse pairs and then bin by time separation $\Delta t$ to create an estimate of the intrinsic correlation coefficient $\rho(\Delta\nu, \Delta t)$ \citep{cordes+04}.

For two spectra $P_1(\nu)$ and $P_2(\nu)$, the expected correlation coefficient is given by,
\begin{equation}
  \rho_{12}(\Delta \nu) = \frac{
    \langle(P_1(\nu)-\mu_1) (P_2(\nu + \Delta\nu)-\mu_2)\rangle}
    {\sigma_1 \sigma_2},
\end{equation}
where $\Delta\nu$ is the offset in frequency, $\mu_1$, $\mu_2$, $\sigma_1^2$ and $\sigma_2^2$ are expectation values for the means and intrinsic variances of $P_1$ and $P_2$, and we use $\langle\dots\rangle$ to indicate the expectation value of the product.  If the giant pulses were effectively delta functions in our band, but affected by the same impulse response function associated with the scintillation, one would expect $\rho=1$ for $\Delta\nu=0$, and a fall-off in frequency and time difference with the appropriate scintillation bandwidth and timescale, approaching 0 at large $\Delta\nu$ and $\Delta t_{12}$.  As noted by \cite{cordes+04}, however, if each pulse consists of multiple shots, the spectra of two pulses will have different structure, and, if still affected by the same impulse response function, one expects a reduced peak, with maximum $\rho\simeq1/3$ (as was indeed observed in their data set).

From observed spectra, one can only estimate the correlation coefficient.  As we show in Appendix~\ref{sec:noise-bias}, if one simply uses the standard equation for the sample correlation coefficient, using the sample mean $m$ and sample variance $s^2$ as estimates of $\mu$ and $\sigma^2$, the result is biased in the presence of background noise, but an unbiased estimate can be made using,
\begin{eqnarray}
  r_{12}(\Delta\nu) &=& \frac{1}{k-1}\sum_{i=1}^{k}
                   \frac{(P_{1}(\nu_i)- m_{1}) (P_{2}(\nu_i+\Delta\nu)-m_{2})}
                   {s_{1} s_{2}}\nonumber\\
  &&\times \left(\frac{m_{1} m_{2}}{(m_{1}-m_{b})(m_{2}-m_{b})} \right),
\end{eqnarray}
where $m_b$ is the mean power in the background.

We create spectra for pulses using the $8\,\mu$s bin centered at each peak, yielding 125\,kHz channels.  To ensure sufficiently reliable correlations between pulse pairs, we limited ourselves for the WSRT data to pulses with S/N $>$ 16, corresponding to S/N $>$ 1 per channel, leaving 6755 main pulses and 650 interpulses.  We use the AR data as a cross-check for possible systematics in the WSRT data, lowering the limit to SN $>$ 10 to have roughly the same sample of pulses, accounting for the ratio of $165/275$ between the two telescope's system temperatures.

\begin{figure*}
\begin{center}
\begin{minipage}[t]{250pt}
\vspace{0pt}
\includegraphics[width=1.0\textwidth]{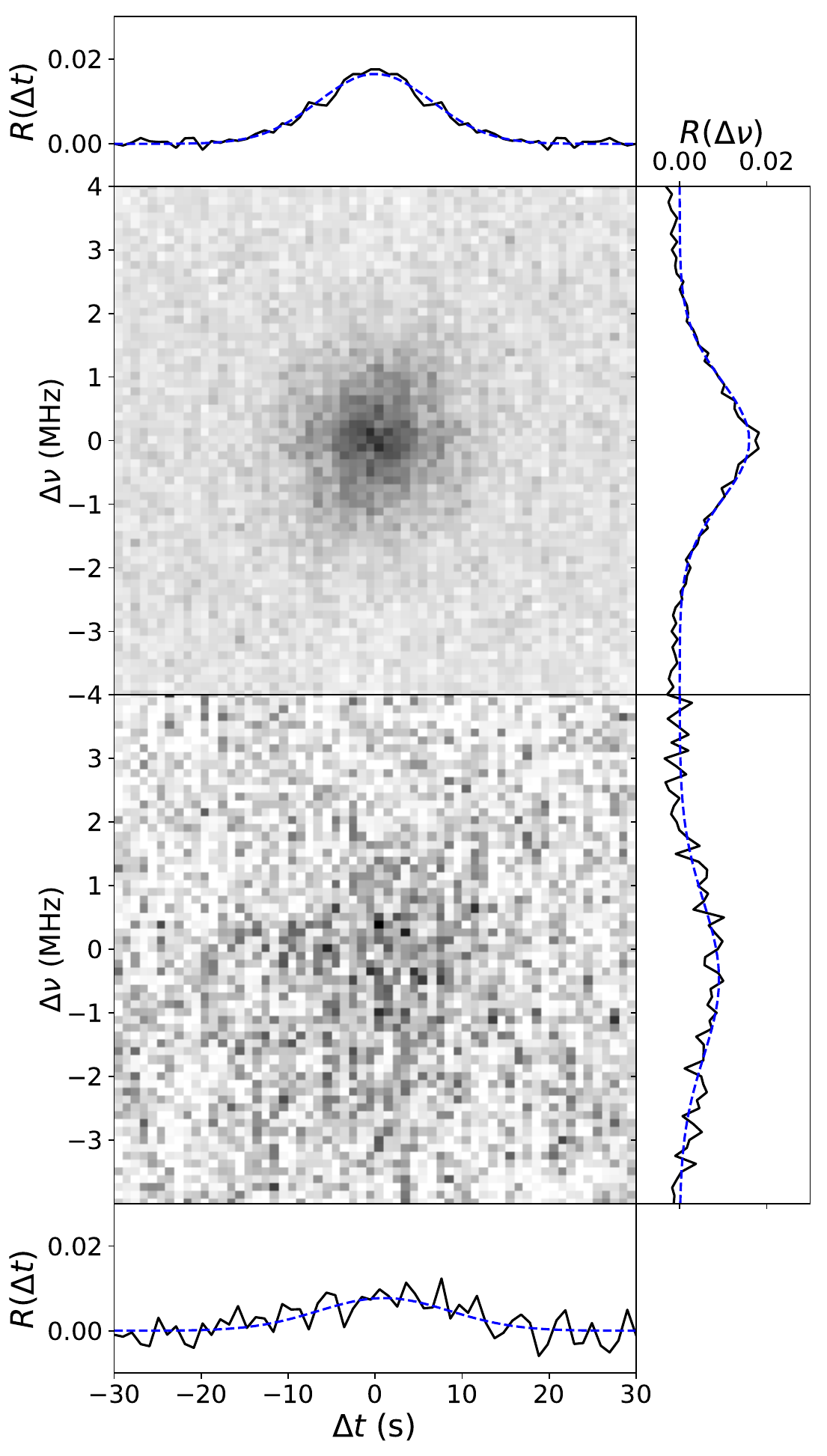}
\end{minipage}
\begin{minipage}[t]{250pt}
\vspace{0pt}
\includegraphics[width=1.0\textwidth]{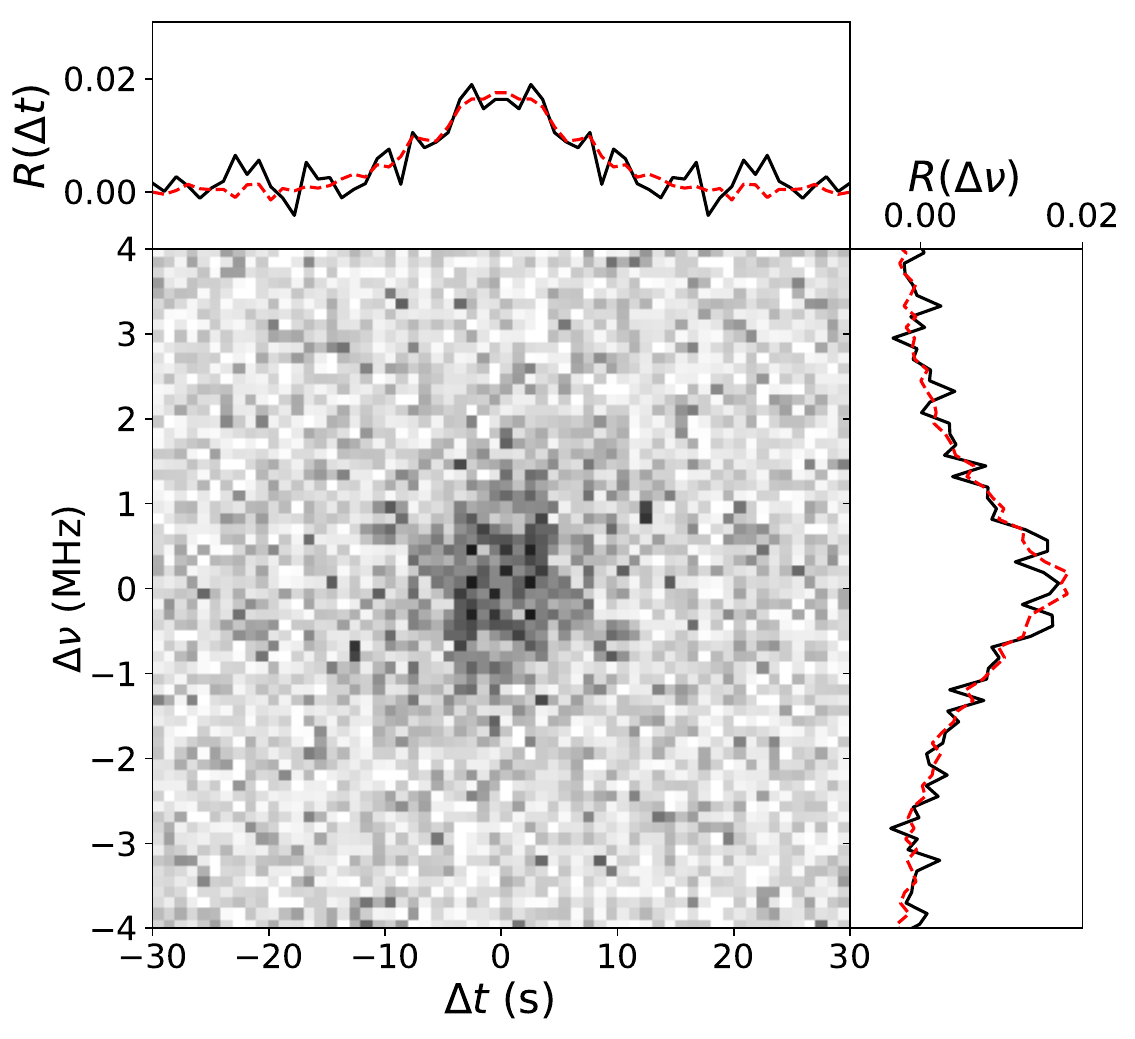}
\caption[Cross-correlations $r(\Delta t, \Delta \nu)$ of giant pulse dynamic spectra]{\textbf{Left:} {\em Images:} Cross-correlations $r(\Delta t, \Delta \nu)$ of pulse dynamic spectra, between giant pulses in the main pulse with themselves ({\em top}) and with giant pulses in the interpulse ({\em bottom}). The correlation between main-pulse giant pulses is symmetric around the origin by construction (i.e., $r(\Delta t, \Delta \nu) = r(-\Delta t, -\Delta \nu)$), but this is not the case for the correlation between interpulse and main pulse.
{\em Side panels:} a 10-bin (-5--5\,s) and 9-bin (-0.5--0.5\,MHz) wide average of correlations through the best fit $\Delta t$, $\Delta \nu$, respectively. Blue dotted lines are the same cuts through the 2D Gaussian fits.
\textbf{Right:} Same as left, but using pulses at AR. The S/N is much lower, owing to the higher $T_{\rm{sys}}$ value at AR (leading to fewer detected pulses, with lower S/N), and the shorter observation time.  The red dotted line is the overlay of the MP-MP correlation at WSRT, showing that the two telescopes give consistent results.
\label{figure:MPIP-ccorr}}
\end{minipage}
\end{center}
\end{figure*}

For each pulse pair, the correlation between their spectra gives $r(\Delta \nu)$ for a single value of $\Delta t$, the time separation of the pulses.  We average these correlated spectra in equally spaced bins of $\Delta t$ to construct our estimate of the correlation function.  We show the result in Fig.~\ref{figure:MPIP-ccorr}, both for correlations between main pulse pairs and for correlations between main pulse and interpulse pairs (there are insufficient giant pulses associated with the interpulse to calculate a meaningful correlation function from those alone).


One sees that the correlations are fairly well defined, and that the correlation of the main pulse with itself is clearly different from that with the interpulse, the latter being broader in both frequency and time, and having lower maximum correlation.  More generally, one sees that the amplitudes of all correlations are surprisingly low.  We investigate the latter further in section \ref{sec:low}, but note here that it is not some systematic product of a given telescope: the results for WSRT and AR are entirely consistent with each other (as expected, as the giant pulses at both telescopes should differ only by noise and by systematics; space-ground VLBI results from \citet{rudnistkii+17} measure the spatial scale of the scintillation pattern of $34000 \pm 9000$\,km during this observation, larger than the Earth's radius).

For more quantitative measures, we fit the correlations with 2D Gaussians with variable amplitude, frequency and time width.  For the correlation between main pulse and interpulse, we additionally allow for offsets in time and frequency $\Delta t_0$ and $\Delta\nu_0$, where the sign convention used is IP - MP (eg. a positive $\Delta t_0$ means the MP precedes the IP).  For the main pulse correlations, we find an amplitude of $1.80\pm0.03\%$ and decorrelation scales of $\nu_{\rm decorr}=1.10\pm0.02\,$MHz in frequency, and $t_{\rm scint}=9.24\pm0.13\,$s in time.\footnote{We adopt the usual convention, defining $\nu_{\rm decorr}$ and $t_{\rm scint}$ as the values where the correlation function drops to $1/2$ and $1/e$ respectively.} The timescale is somewhat shorter than the value of $25\pm5\,$s found at $1.475\,$GHz by \citet{cordes+04}, and the difference in observing frequency does not account for the difference (for $t_{\rm scint}\propto\nu$, our measurement corresponds to $8.17\pm0.12\,$s at $1.475\,$GHz). Differences are expected for observations at different epochs, however, as the scattering in the nebula is highly variable (\citealt{rankin+73, lyne+75, isaacman+77, rudnistkii+17}, and often showing ``echoes'', e.g., \citealt{backer+00, lyne+01, driessen+19}).

Our fits to the main pulse to interpulse correlations confirm the qualitative impression from Fig.~\ref{figure:MPIP-ccorr}, that compared to the main pulse to main pulse correlations they are weaker and broader in both frequency and time:
the measured amplitude is 0.97$\pm$0.07\%, $t_{\rm scint} = 10.7\pm0.8$\,s, $\nu_{\rm decorr} = 1.44\pm0.10$\,MHz.  We also find marginally significant time and frequency offsets, of $\Delta t_0 = 1.02\pm0.54$\,s and $\Delta\nu_0 = -0.34 \pm 0.09$\,MHz.

\begin{figure*}
\begin{center}
\includegraphics[width=1.0\textwidth]{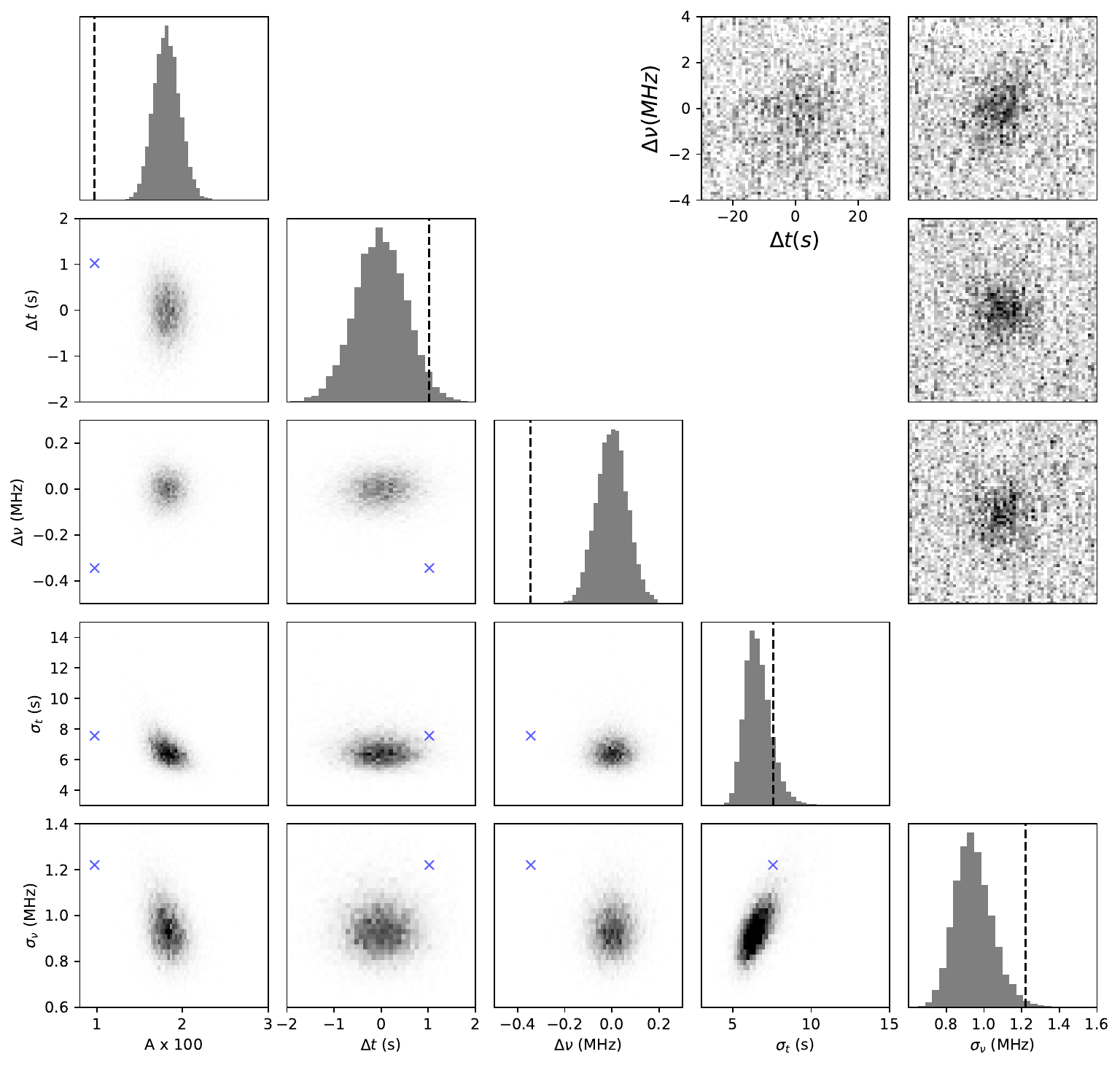}
\caption[Corner plot of the best-fit parameters of the simulated interpulse to main-pulse correlation function.]{{\em Bottom-left:} Corner plot of the best-fit parameters of the simulated interpulse to main-pulse correlation function, obtained by fitting a two-dimensional Gaussian.  Simulated correlation functions are constructed from randomly drawn sets of giant pulses from the main pulse (with the same sample size as that available for the interpulse), correlated with the full main pulse sample. The dotted lines, and the blue crosses show the best fit to the actual MP-IP correlation.
{\em Top-right:} The MP-IP correlation, and 3 example simulated MP-IP plots for comparison.
\label{figure:sims}}
\end{center}
\end{figure*}

To try to quantify the significance of these differences, we use simulated cross-correlations. For these, since we have many more giant pulses during the main pulse than the interpulse, we simply take 650 random main pulses (the number of interpulses above $16 \sigma$) and correlate these with the other 6755 main pulses, without correlating identical pulses.  We repeat this 10000 times, and fit each subset with a 2D Gaussian, allowing for offsets in time and frequency $\Delta t_0$ and $\Delta\nu_0$.  Comparing these with the value fit to the interpulse to main-pulse correlations (see Fig.~\ref{figure:sims}), the differences appear significant: none of the simulated data sets have as small an amplitude, or larger frequency offset $\Delta \nu_0$, while only relatively small numbers have larger time offset $\Delta t_0$ or wider frequency or time widths.

\subsection{Comparison to previous work on the same dataset}

The same data analysed in this paper were studied in \citet{rudnistkii+17} and \citet{popov+17}, who derive a de-correlation bandwidth of $279.2 \pm 34.4$\,kHz, and $320$\,kHz, respectively. These values differ significantly from our value of $\nu_{\rm decorr}=1.10\pm0.02\,$MHz, so here we further investigate the origin of these differences.

Our methods differ to those of \citet{rudnistkii+17} and \citet{popov+17}; the crucial difference being that they auto-correlate individual giant pulses (between left and right circular polarization, to reduce intrinsic structure correlating), while we correlate pulse pairs.
For a direct comparison, we try to follow the steps of \citet{popov+17}, adopting their cutoff of SN $> 22$, correlating left and right circular polarizations of each giant pulse and fitting a single exponential.  From this, we measure $\nu_{\rm decorr}=0.39$\,MHz, much closer to their value of $\nu_{\rm decorr}=0.32$\,MHz. However, we find that a two-exponential fit is a much better fit to the data, giving two distinct scales of $\nu_{\rm decorr, 1}=1.0$\,MHz, $\nu_{\rm decorr, 2}=0.19$\,MHz. This is consistent with our results if the small bandwidth $\nu_{\rm decorr}$ is caused by intrinsic pulse structure (correlating only within a single pulse's spectrum), and the wide bandwidth $\nu_{\rm decorr}$ is the scintillation bandwidth (correlating between pulse pairs within $t_{\rm scint}$).

Additionally, \citet{rudnistkii+17} derive a scintillation timescale of $22.4 \pm 6.1$, larger than our value by a factor of 2.  Their timescale is derived in a different way, where they use their measured scintillation bandwidth (described above), and angular size $\theta$ of the scattering screen, derived directly from their VLBI correlation.  Using the known velocity of the Crab, and an assumed isotropic scattering screen, gives a timescale estimate.  Given the difference in our methods, and the fact that the angular broadening likely arises in the interstellar medium, rather than the nebula, we are not worried about the differences in these values.

\subsection{Secondary Spectra}

\begin{figure*}
\begin{center}
\includegraphics[trim=0.cm 0cm 0.cm 0cm, clip=true, width=\textwidth]{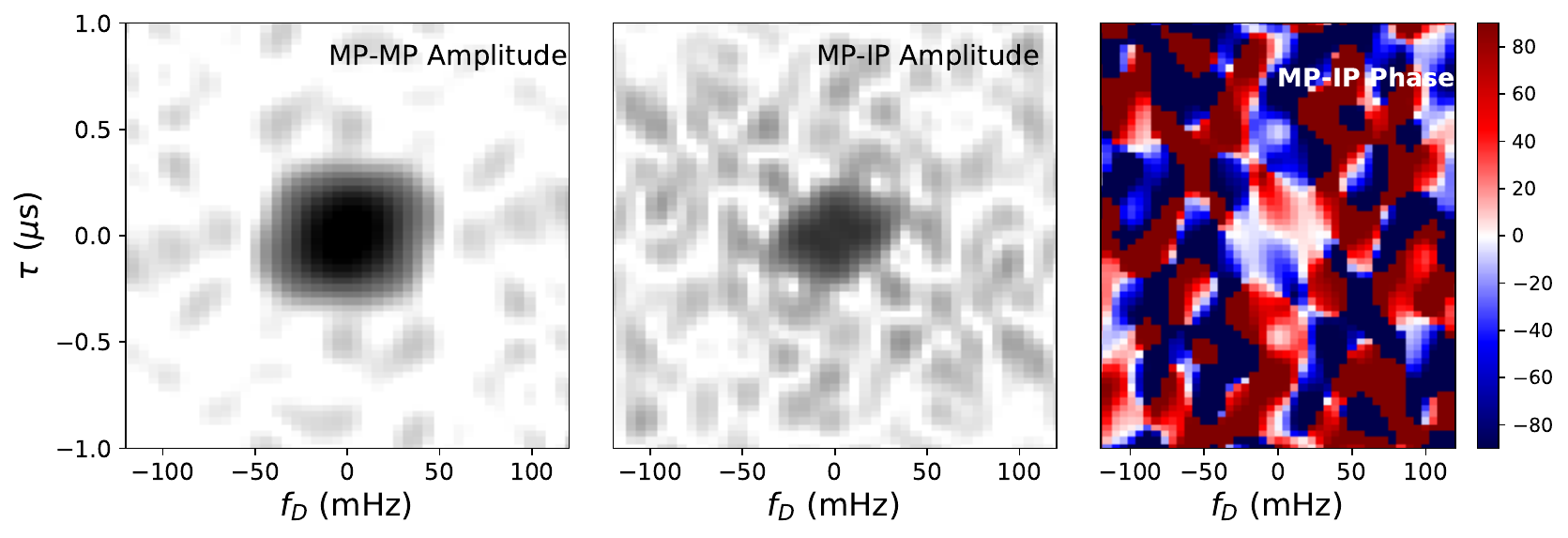}
\caption[secondary spectrum of main pulse, and secondary cross-spectra of the main pulse - interpulse correlation]{{\em Left:} secondary spectrum of the main pulse. {\em Middle and Right:} Amplitude and phase of the main pulse - interpulse cross-spectrum. Note the amplitude and phase are point symmetric, and point anti-symmetric, by definition. There appears to be a phase gradient in $\tau$, reflecting the offset in frequency in the MP-IP correlation. Similarly, the possible phase gradient along~$f_{D}$ reflects the marginally significant offset in time between the main pulse and interpulse.
\label{figure:CrabSS} }
\end{center}
\end{figure*}

Pulsar scintillation is often best studied in terms of its conjugate variables $\tau$ and $f_{D}$, through their secondary spectrum $A(\tau, f_{D}) = \tilde{I}_{1}(\tau, f_{D}) \tilde{I}_{2}^{*}(\tau, f_{D})$ (e.g. \citealt{stinebring+01, brisken+10}).  The secondary spectrum is simply the Fourier transform of the correlation function $r(\Delta \nu, \Delta t) = I_{1}(\nu, t) \circledast I_{2}(\nu, t)$.  For the MP-MP correlation, $A(\tau, f_{D})$ is purely real, but in the MP-IP correlation, any time or frequency offsets in the correlation function will manifest as phase gradients in $f_{D}$ or $\tau$, respectively.  We show the secondary spectra for both correlations, after padding by 60 zero bins in time, in Fig.~\ref{figure:CrabSS}.

The MP-IP secondary spectrum is dominated by a phase gradient in $\tau$, arising from the frequency offset in the correlation function. Removing a linear phase gradient in $\tau$ shows a marginally significant phase gradient in $f_{D}$. If the screen was one-dimensional, and the main pulse and interpulse emission locations were offset, there would be a phase gradient in $f_{D}$, independent of $\tau$.

\section{Ramifications}
\label{sect:ramifications}

\subsection{The Surprisingly low Correlation Coefficient}
\label{sec:low}

Giant pulses are on average a few $\mu$s in duration, comprised of many smaller, unresolved ``nanoshots'' (e.g. \citealt{hankins+07}). However, if all nanoshots originate from the same projected physical location, they should all be imparted with the same impulse response function; an identical signal would correlate perfectly, and a signal with many random polarized shots with the same impulse response should correlate no worse than 1/3 (\citealt{cordes+04}, Appendix ~\ref{sec:noise-bias}).

The observed $\sim 2\%$ correlation between main pulses is well below the expectation of $1/3$.
This could be explained if individual pulses come from small parts of the full extended emission region, which is larger than the resolution of the scattering screen
(discussed in the following section).
Under this explanation, the correlation should decrease even further during times of higher nebular scattering - this is something we investigate further in Lin et al.\ (2020, in preparation).

\subsection{Spatial Resolution of Scattering Screen}
\label{sect:resolution}

The size and location of the scattering screen is not precisely known,
but a model in which the majority of the temporal scattering occurs in the Crab nebula is favoured by VLBI measurements showing the visibility amplitude is constant through the scattering tail, and independent of the scattering time (\citealt{vandenberg+76}, \citealt{vandenberg+76b}) as well as by the short scintillation timescale (\citealt{cordes+04}).

Since scattering requires relatively large differences in (electron) density, it is very unlikely to happen inside the pulsar-wind filled interior of the Crab nebula, which must have very low density.  For a reasonable bulk magnetic field of $10^{-4}$ G, the emitting radio electrons are very relativistic, with $\gamma \sim 10^{3}$. The radio emitting electrons have a density of $n_e \approx 10^{-5}{\rm\,cm^{-3}}$ \citep{shklovsky57}, implying that the refractive index deviates from unity by a tiny amount,
\begin{equation}
\Delta n \approx \left(\frac{\nu_p}{\nu} \right)^2 \sim 10^{-18},
\end{equation}
where $\nu_p = (e^2 n_e / \pi \gamma m_e)^{1/2}$ is the plasma frequency, and $\nu$ the observed radio frequency.

Instead, the only plausible location for the temporal scattering is in the optically emitting filaments in the Crab Nebula, which have $n_{e} \sim 1000{\rm\,cm^{-3}}$ \citep{osterbrock57}.
These filaments develop because as the pulsar wind pushes on the shell material, the contact discontinuity accelerates, leading to the RT instability \citep{chevalier+77, porth+14}.

With 3-dimensional models fit to optical spectroscopic data of the Crab Nebula, \citet{lawrence+95} find the filaments reside
conservatively in the range $\sim 0.5--2.0$\,pc
when using a nominal pulsar distance of 2 kpc.

The scattering causes a geometric time delay given by,
\begin{equation}
  \tau = \frac{\theta^2 d_{\rm eff}}{2c},
  \qquad\mbox{with}\qquad
  d_{\rm eff} = \frac{d_{\rm psr}d_{\rm lens}}{d_{\rm psr} - d_{\rm lens}},
\end{equation}
where $\theta$ is the angle the screen extends to as seen from Earth, and $d_{\rm psr}$ an $d_{\rm lens}$ are the distances to the pulsar and the screen, respectively.

The scattering screen can be seen as a lens, with physical size $D=\theta d_{\rm lens}$ and corresponding angular resolution $\lambda/D$, giving a physical resolution at the pulsar of $\Delta x = (d_{\rm psr} - d_{\rm lens}) \lambda / \theta d_{\rm lens}$, or, in terms of the scattering time $\tau$,
\begin{equation}
  \Delta x = \frac{\lambda}{\sqrt{2} \pi}\left(\frac{d_{\rm psr} - d_{\rm lens}}{2c\tau}
                            \frac{d_{\rm psr}}{d_{\rm lens}}
                    \right)^{1/2}.
\end{equation}
The prefactor $1 / \sqrt{2} \pi$ is model dependent, coming from using a square-law  ($\alpha = 2$) phase-structure function \citep{cordes+98}.

If we were to infer the scattering time from the scintillation bandwidth, we would find $\tau_{\rm scint} = 1/2\pi\Delta\nu \simeq 160{\rm\,ns}$, using $\Delta\nu \approx 1\ \rm{MHz}$.  However, this is lower than the apparent $\sim\!1\,\mu$s scattering seen in Figure \ref{fig:pulses}, and lower than measurements of the scattering  at this epoch of $\tau(600\,\rm{MHz}) \simeq 0.1\,$ms \citep{mckee+18}, or $\tau(350\,\rm{MHz}) \simeq 0.6\,$ms \citep{driessen+19}, which would correspond to $\tau \simeq 1.1-1.7 \mu$s  when scaled by $\tau \propto \nu^{-4}$ to our observing frequency.  In \citet{gwinn+98}, it is noted that the relation $\tau = 1/2\pi\Delta\nu$ may underestimate $\tau$ in the case of an extended, resolved emission region as
\begin{equation}
    \tau = \frac{\sqrt{1+4\sigma_{1}^{2}} }{2\pi \Delta\nu},
    \label{eq:gwinn}
\end{equation}
where $\sigma_{1}$ is the size of the emission region in units of the lens resolution.
In this picture, the scintillation timescale would depend on the lens resolution and the emission region size.

Using $\tau\simeq1{\rm\,\mu s}$ and $d_{\rm psr}-d_{\rm lens}\simeq1.0\,$pc,
then $\Delta x\simeq290\,$km (for the full range of allowed distances, $205\lesssim \Delta x\lesssim410{\rm\,km}$).
Thus, the resolution of the scattering screen is smaller than the light-cylinder radius of the Crab pulsar, $R_{\rm LC}\equiv cP/2\pi=1600\,$km.

Additionally, a nominal time offset between the main pulse and interpulse of $\sim 1-2\,$s is $\sim 10-20 \%$ of the scintillation timescale, which would suggest that the emission locations are separated by hundreds of km.
We could turn a measured time offset into a physical separation given a relative velocity between the pulsar and the screen.  Unfortunately, this is not known, though we can set limits from the proper motion.
The proper motion of the Crab pulsar relative to its local standard of rest is measured to be $12.5\pm2.0{\rm\,mas/yr}$ in direction $290\pm9{\rm\,deg}$ (east of north \citealt{kaplan+08}), where the uncertainties attempt to account for the uncertainty in the velocity of its progenitor, and, therewith, of the nebular material. At an assumed distance of 2\,kpc, the implied relative velocity of the pulsar is $\sim\!120{\rm\,km/s}$, and non-radial motions in the filaments can be up to $\sim\!70\,$km/s \citep{backer+00}.  A 1-2\,s time delay between pulse components would then suggest a projected separation between the interpulse and main pulse emission regions of $ \sim 50 - 400$\,km.

As mentioned above, \citet{cordes+04} argue that the short scintillation timescale suggests a nebular origin of the observed scintillation. Here we outline the argument using our measured values.
The scintillation timescale is roughly the time it takes for the extended emission region to traverse a resolution element of the scattering screen; using the above resolution and proper motion gives an estimate of the timescale of
$ \sqrt{\sigma_{1}^{2}+1}\Delta x / v_{\rm pm} \sim 5.5-11$\,s, consistent our observed time of $9.24\pm0.13\,s$.

Scintillation in the interstellar screen for our given scintillation bandwidth would result in much larger resolution elements (for a screen halfway to the pulsar, at $d_{\rm psr}-d_{\rm lens}\simeq1{\rm\,kpc}$, greater by a factor $\sim \sqrt{{\rm 1\,kpc/1\,pc}} \sim 30$), and scintillation on several minute timescales, more typical of interstellar scintillation in this frequency range.

If we assume pulses occur at random position within an extended region, we may also estimate the average expected correlation. To test this, we simulated 500 pulses with position drawn at random from a 2D Gaussian with $\sigma_{xy} = \sigma_{1} \Delta x$.  The correlation coefficient between each pair of pulses is estimated as $r_{ij} = C e^{-|\vec{x_{ij}}|^{2} / (\Delta x) }$, where $|\vec{x_{ij}}|$ is the projected position difference between each pulse pair,
and $C < 1$ is an unknown constant which depends on both the intrinsic spectral structure in the pulses (eg. \citealt{cordes+04} and the appendix), and whether the separate components forming giant pulses are resolved.
Assuming an isotropic 2D screen, or a 1D screen, and assuming $C=1/3$ (ie. that individual giant pulses are unresolved, which may not be a good assumption), gives estimates of the expected correlation coefficient of $\langle r_{1D} \rangle \simeq 0.071$,  $\langle r_{2D} \rangle \simeq 0.016$ respectively, on the same order as our observed average correlation coefficient.  We find the assumed picture of giant pulses occurring from an extended region of of $\sim 1000\,$km to give a consistent result, being broadly in agreement with the observed scintillation bandwidth, scintillation timescale, and low correlation coefficient. This picture will be expanded in more detail in Lin et al.\ (in prep.).

The picture we find above differs from \citet{cordes+04}, who find the spectra of nearby pulses at $1.48\,$GHz and $2.33\,$GHz to correlate at a value of $\sim 1/3$.  They find values of the scintillation bandwidth and timescale at $2.33\,$GHz of $\Delta \nu_{s}=2.3\pm0.4\,$MHz, and $\Delta t_{s}=35\pm5\,$s respectively, which scaled to $1.68\,$GHz gives $\Delta \nu = 0.6\pm0.1\,$MHz, and $\Delta t = 25\pm4\,$s (\citealt{cordes+04} also consistently find $\Delta \nu_{s} < 0.8\,$MHz, $\Delta t_{s}=25\pm5\,$s at $1.48\,$GHz).  They face a similar inconsistency between the measured scattering time ($\sim 0.1\,$ms at $600\,$MHz, which imples $\sim\!1.7\,\mu$s at $1.67\,$GHz, \citealt{mckee+18}) and the inferred scattering time from $1/2\pi \Delta \nu \approx 250\,$ns.  However, the scintillation time they measure is much larger than ours, implying that the dominant screen must be at a further distance from the pulsar, or anisotropic and oriented such that there is poorer resolution along the direction of the relative velocity between the pulsar and the screen.

\subsection{Fully Quantifying Emission Sizes and Separations}

A major uncertainty in our above estimates
is the geometry of the lens.  From studies of the scintillation in other pulsars, the scattering screens in the interstellar medium are known to be highly anisotropic, as demonstrated most dramatically by the VLBI observations of \citet{brisken+10}.  If the same holds for the nebular scattering screens, this implies that our resolution elements are similarly anisotropic.  Since the orientation relative to the proper motion is unknown, the physical distance between the main and interpulse regions could be either smaller or larger than our estimate above.  Since the scattering varies with time, it may be possible to average out these effects.

With a perfectly 1D scattering screen, it is difficult to produce both a time and frequency offset, as there would necessarily be some position where the main pulse and interpulse pass through the same position along the screen's axis.  One possible way to induce a frequency offset would be a spatial gradient of the column density (or ``prism'') on the scale of separation between emission regions; Our frequency offset of $\sim0.3$\,MHz could be explained by a DM gradient of $\Delta \mathrm{DM} / \mathrm{DM} \sim 0.02\%$ over $\sim 1000$\,km.  The DM variations of the Crab have not been probed on such small spatial scales, although it varies by considerably more than this on longer timescales (ie. larger spatial scales, eg. \citealt{mckee+18}).
For two spatially separated emission components, \citet{ravi+18} find that a two-dimensional screen can produce both a time and frequency shift, over a short timescale.
The two-dimensionality of the screen, or the effect of multiple scattering screens, may need to be considered.


Furthermore, all values relating to the scattering screen include the uncertain distance to the Crab pulsar, suggesting that a parallax distance would improve our constraints.  In addition, the rough localization of the scattering in the filaments is based on physical arguments; the results would be greatly improved through a direct measurement.

The distance to the screen(s) can be constrained through VLBI and through scintillation measurements across frequency. As the spatial broadening of the Crab is dominated by the interstellar screen, rather than the Nebula, VLBI at space-ground baselines \citep{rudnitskii+16} or at low frequencies \citep{vandenberg+76b}
can help constrain the angular size of the scattering in the interstellar medium.
This in turn can constrain the size of the nebular screen.  The visibility amplitudes will only decrease below 1 when the scattered image of the pulsar is not point-like to the interstellar screen.
Time-resolved visibilities throughout the rise time of scattered pulses may then elucidate the neublar scale; by increasing in time delay, one increases in angular size, and thus resolution, so one may observe the transition point beyond which the nebular screen becomes resolved.
In addition, the interstellar screen will scintillate only when it does not resolve the nebular screen (the same argument has been made for scintillation in FRBs, \citealt{masui+15}).
The transition frequency for the two scintillation bandwidths to become apparent in the spectra could give a size measurement of the nebular screen.

Applying this same analysis across different frequencies, or in times of different scattering in the nebula will also help to quantify both the separation of the main pulse and interpulse, and the size of the emitting regions of both components. The correlation function of spectra is a crude measurement - it is fourth order in the electric field, and the scintillation pattern is contaminated with intrinsic pulse substructure.  A much cleaner measurement can hopefully be made in the regime where the duration of giant pulses is much less than the scattering time, akin to the coherent method of de-scattering pulses in \citet{main+17}.

As discussed in section \ref{sect:resolution}, we associate the scattering screen with the filaments in the pulsar wind nebula \citep{porth+14}. These filaments appear from the Rayleigh-Taylor instability, when the pulsar wind pushes and accelerates freely expanding envelope. This stage terminates after few thousand years when the reverse shock from the interaction between the supernova remnant at the ISM reaches the pulsar wind nebula (\citealt{gelfand+09}, see review by \citealt{slane17}). Thus we expect such special scattering environments to be specific for pulsar wind nebulae during a fairly short period - sufficiently young for the reverse shock not to reach the pulsar wind nebula, but sufficiently advanced to have RT-induced filaments.

\subsection*{Acknowledgements}

We thank the anonymous referee, whose comments greatly improved the draft.
We thank Judy Xu who attempted the initial 1D correlation function of giant pulses.  These data were taken as part of a RadioAstron observing campaign. The RadioAstron project is led by the Astro Space Center of the Lebedev Physical Institute of the Russian Academy of Sciences and the Lavochkin Scientific and Production Association under a contract with the State Space Corporation ROSCOSMOS, in collaboration with partner organizations in Russia and other countries.

\newpage

\appendix
\section*{Correcting Noise Biases in the Correlation Coefficient}
\label{sec:noise-bias}

The intrinsic correlation coefficient between two pulse spectra $P_{1,2}(\nu)$ can be generally defined as,
\begin{equation}
\rho(P_{1}, P_{2}) = \frac{\langle (P_{1}(\nu)- \mu_{1}) (P_{2}(\nu)-\mu_{2}) \rangle}{\sigma_{1} \sigma_{2}},
\label{eq:cc}
\end{equation}
where $\langle\dots\rangle$ indicates the expectation value for an average over frequency, and $\mu$ and $\sigma^2$ are expectation values of the mean and variance, respectively.  With this definition, one will have $\rho=1$ for two pulses with identical frequency structure.

Typically, as an estimate of $\rho$, one uses the sample correlation coefficient,
\begin{equation}
  r(P_1, P_2) = \frac{1}{k-1}\sum_{i=1}^k
  \frac{(P_1(\nu_i) - m_1) (P_2(\nu_i) - m_2)}
  {s_1 s_2},
\label{eq:sample_cc}
\end{equation}
where $k$ is the number of frequency bins and $m_p$ and $s_p^2$ are the usual sample mean and variance.  In the presence of noise, subtracting $m$ leaves the nominator unbiased, but $s^2$ will be systematically higher than $\sigma^2$, and thus $r$ will be biased low.  For normally distributed data, one could approximately correct with $s_{\rm int}^2=s_p^2-s_n^2$, but this does not hold for our case of power spectra.

Here, we derive an expression valid for our case, where we wish to ensure that $\langle r\rangle=1$ for two pulses that are sufficiently short that we can approximate them as delta functions, and that are affected by the interstellar medium the same way, i.e., have the same impulse response function $g(t)$.  In that case, the measured electric field of a giant pulse is,
\begin{equation}
E_p(\nu) = A_p g(\nu) + n(\nu),
\end{equation}
where $A_p$ is the amplitude of the pulse's delta function in the Fourier domain, and $g(\nu)$ and $n(\nu)$ are the Fourier transforms of the impulse response function and the measurement noise, respectively. The measured intensity is then
\begin{equation}
P_p(\nu) = E_p^2(\nu) = A_p^2 g^2(\nu) + n^2(\nu) + 2A_p|g(\nu)||n(\nu)|\cos(\Delta \phi(\nu)),
\end{equation}
where $\Delta \phi(\nu)$ is the phase difference between $n(\nu)$ and $g(\nu)$, and where squares are of the absolute values.

The expectation value for the average is,
\begin{equation}
\mu_p = \langle P_p \rangle = A_p^2 \langle g^2 \rangle + \langle n^2 \rangle,
\end{equation}
where we have dropped the dependencies on frequency for brevity, and used that the cross term averages to zero since $\langle\cos(\Delta\phi)\rangle=0$. Hence, the expectation value for the variance is,
\begin{equation}
\sigma_p^2 = A_p^4\left[\langle g^4\rangle-\langle g^2\rangle^2\right]
      + \langle n^4\rangle - \langle n^2\rangle^2
      + 4A_p^2 \langle g^2n^2\cos^2(\Delta\phi)\rangle,
\end{equation}
where we have again omitted terms that average to zero. The last term does not average to zero because of the squaring: it reduces to $2A_p^2\langle g^2\rangle\langle n^2\rangle$, since $g$ and $n$ are independent and $\langle\cos^2(\Delta\phi)\rangle=1/2$.

For two pulses differing only by noise, the expectation value for the numerator of $r$ is
\begin{equation}
\langle (P_{1}(\nu)- \mu_{1}) (P_{2}(\nu)-\mu_{2}) \rangle = A_1^2 A_2^2 \left[\langle g^4\rangle-\langle g^2\rangle^2\right].
\end{equation}
Thus, for an unbiased estimate of $r$, we need to estimate $\sigma_p=A_p^2\left[\langle g^4\rangle-\langle g^2\rangle^2\right]^{1/2}$. We can do this by also measuring the properties of the background, which, if it is dominated by measurement noise with the same properties as the pulse, has $\mu_b=\langle n^2\rangle$ and $s^2_b=\langle n^4\rangle - \langle n^2\rangle^2$ (this will underestimate the noise if the pulse is strong enough to raise the system temperature, although in that case, the noise has only a small contribution to the variance of the pulse, and is negligible in computing $r$.
With this, it follows that to make estimates of $r$ free of noise bias, we should use,
\begin{equation}
s_{\rm int}^2 = s_p^2 - s_b^2 - 2(\mu_p - \mu_b)\mu_b.
\end{equation}
This is a noisy quantity, however, and simply replacing the measured variance with this value will lead to some measured values of $s_{\rm int}\sim0$, and thus diverging correlations.

Instead of correcting the sample variances in the denominator of the sample correlation coefficient, one can use the properties of the pulse to estimate a correction of the sample correlation coefficient itself.  Assuming the impulse response function $g(\nu)$ is approximately normally distributed, the power spectrum $|g(\nu)|^{2}$ will distributed roughly as a $\chi^{2}$ distribution with two degrees of freedom, with $s_{p}^{2} \simeq m_{p}^{2}$. Using this, our unbiased estimate of the variance simplifies to $s_{\rm int}^{2} = (m_p - m_b)^{2}$, which uses the well-measured mean of both the pulse and the background (in our case, after bandpass calibration described in section \ref{sect:data}, the mean and standard deviation of the background are unity).

We should not use the above estimate directly in the denominator of the standard correlation coefficient in Eq.~\ref{eq:cc}, as at high S/N, this unnecessarily introduces extra variance.  Instead, we can ensure an unbiased, noise-corrected estimate of the correlation that works at both low and high S/N by writing it as:
\begin{equation}
r(P_{1}(\nu), P_{2}(\nu)) = \frac{\langle (P_{1}(\nu)- m_{1}) (P_{2}(\nu)-m_{2}) \rangle}{s_{1} s_{2}} \left(\frac{m_{1} m_{2}}{(m_{1}-m_{b})(m_{2}-m_{b})} \right).
\end{equation}

\begin{figure*}
\begin{center}
\includegraphics[ width=0.9\textwidth]{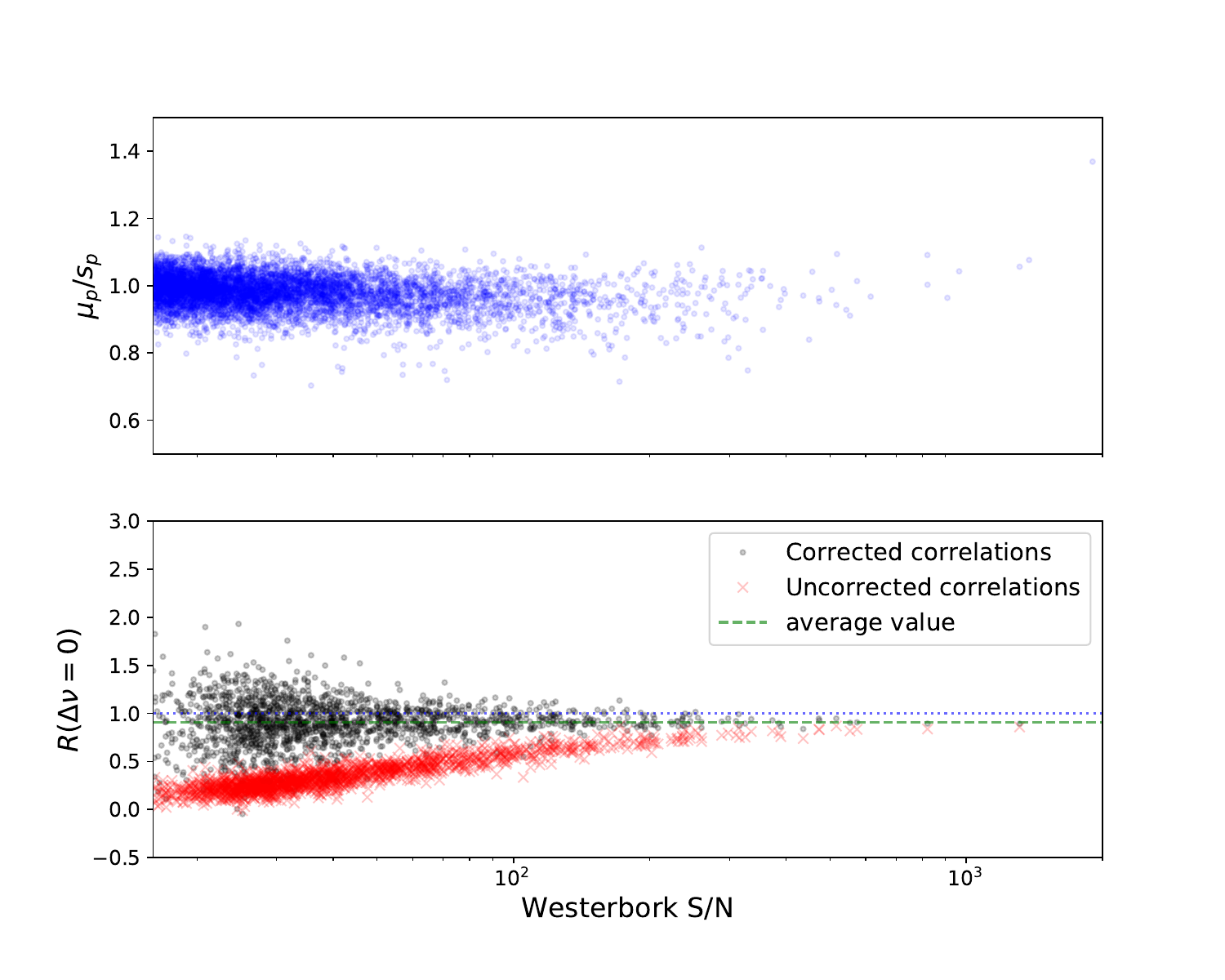}
\caption[Westerbork to Arecibo correlation]{
{\em Top:} Ratio of the mean and standard deviation of every giant pulse spectrum.  It averages at 1 independently of S/N, showing that it is reasable to use this as an assumption for correcting our correlation coefficients.
{\em Bottom:} Sample correlation coefficients of giant pulse spectra for each pulse detected above $16 \sigma$ at both WSRT and AR, ordered by increasing S/N.  For the black points, the noise correction described in the appendix has been applied, while for the red points no correction has been applied (i.e., the standard formula for the sample correlation coefficient has been used).
\label{figure:wbar}}
\end{center}
\end{figure*}

To test our correlation correction, we first verify our underlying assumption, that the mean and standard deviation are equal, by taking the ratio of the measured values for all pulses. The result is shown in the top panel of Figure~\ref{figure:wbar}: one sees that the ratio is around unity, except at the highest S/N where saturation biases the noise low.  Next, we correlated the spectra of all overlapping pulses between WSRT and AR with S/N $>$ 16 at both telescopes, shown in the bottom panel of Figure~\ref{figure:wbar}.  Since these are the same pulses, but observed at two different telescopes and thus with different noise, we expect values to scatter around unity.  Indeed, using the full noise correction, we find that the sample correlations are around a value close to unity, of $\sim\!90\%$, independently of S/N (the difference from 100\% likely reflects remaining differences in bandpass, etc.; the baseline is too short for interstellar scintillation to differ).  Without the correction, the sample correlation coefficient is always less than $100\%$, and decreases with decreasing S/N.

For a further test, we use simulated giant pulses. To begin, pulses are simulated as identical delta function giant pulses with the same impulse response function but different noise, in the manner described in \citet{main+17}. We find that using the above estimates, the correlation coefficients between these pulses indeed average to unity.  Trying a slightly more realistic simulation, forming giant pulses with $N$ fully polarized shots, with random amplitudes (drawn from a normal distribution) and random phases, the correlation decreases, saturating at $r = 1/3$ for large numbers ($N\gtrsim 10$), in line what is expected from the derivation in \citet{cordes+04}.

Finally, all of the above is for correlations of the power spectra of a single polarization, and throughout the paper, we correlate each polarization separately, then average the two values.  For completeness, we note that if one were to use the total intensity  $I = P_{L} + P_{R} = P_{X} + P_{Y}$, under the assumption that the noise is not correlated between the polarizations, the expectation value for the standard deviation is
\begin{equation}
\sigma_I^2 = A_p^4\left[\langle g^4\rangle-\langle g^2\rangle^2\right]
      + \langle n^4\rangle - \langle n^2\rangle^2
      + 2A_p^2 \langle g^2n^2\cos^2(\Delta\phi)\rangle,
\end{equation}
and a noise corrected estimate can be made with,
\begin{equation}
s_{\rm int}^2 = s_I^2 - s_b^2 - (\mu_I - \mu_b)\mu_b,
\end{equation}
with the cross-term differing by a factor of 2 from the single-polarization case.  Adding more samples with independent noise, the cross-term further diminishes, and can be treated as Gaussian in the limit of large N.  We do not use these estimates, however, since in the case that the giant pulses are not single delta functions, the estimates start to depend on the degree of polarization \citep{cordes+04}, which is a complication we rather avoid.

\bibliographystyle{apj}
\bibliography{CrabWB}

\begin{thebibliography}{}
\expandafter\ifx\csname natexlab\endcsname\relax\def\natexlab#1{#1}\fi

\bibitem[{{Abdo} {et~al.}(2010){Abdo}, {Ackermann}, {Ajello}, {Atwood},
  {Axelsson}, {Baldini}, {Ballet}, {Barbiellini}, {Baring}, {Bastieri},
  {Bechtol}, {Bellazzini}, {Berenji}, {Blandford}, {Bloom}, {Bonamente},
  {Borgland}, {Bregeon}, {Brez}, {Brigida}, {Bruel}, {Burnett}, {Caliandro},
  {Cameron}, {Camilo}, {Caraveo}, {Casandjian}, {Cecchi}, {{\c C}elik},
  {Chekhtman}, {Cheung}, {Chiang}, {Ciprini}, {Claus}, {Cognard},
  {Cohen-Tanugi}, {Cominsky}, {Conrad}, {Dermer}, {de Angelis}, {de Luca}, {de
  Palma}, {Digel}, {Silva}, {Drell}, {Dubois}, {Dumora}, {Espinoza}, {Farnier},
  {Favuzzi}, {Fegan}, {Ferrara}, {Focke}, {Frailis}, {Freire}, {Fukazawa},
  {Funk}, {Fusco}, {Gargano}, {Gasparrini}, {Gehrels}, {Germani}, {Giavitto},
  {Giebels}, {Giglietto}, {Giordano}, {Glanzman}, {Godfrey}, {Grenier},
  {Grondin}, {Grove}, {Guillemot}, {Guiriec}, {Hanabata}, {Harding},
  {Hayashida}, {Hays}, {Hughes}, {J{\'o}hannesson}, {Johnson}, {Johnson},
  {Johnson}, {Johnson}, {Johnston}, {Kamae}, {Katagiri}, {Kataoka}, {Kawai},
  {Kerr}, {Kn{\"o}dlseder}, {Kocian}, {Kramer}, {Kuehn}, {Kuss}, {Lande},
  {Latronico}, {Lee}, {Lemoine-Goumard}, {Longo}, {Loparco}, {Lott},
  {Lovellette}, {Lubrano}, {Lyne}, {Makeev}, {Marelli}, {Mazziotta}, {McEnery},
  {Meurer}, {Michelson}, {Mitthumsiri}, {Mizuno}, {Moiseev}, {Monte},
  {Monzani}, {Moretti}, {Morselli}, {Moskalenko}, {Murgia}, {Nakamori},
  {Nolan}, {Norris}, {Noutsos}, {Nuss}, {Ohsugi}, {Omodei}, {Orlando}, {Ormes},
  {Ozaki}, {Paneque}, {Panetta}, {Parent}, {Pelassa}, {Pepe}, {Pesce-Rollins},
  {Pierbattista}, {Piron}, {Porter}, {Rain{\`o}}, {Rando}, {Ray}, {Razzano},
  {Reimer}, {Reimer}, {Reposeur}, {Ritz}, {Rochester}, {Rodriguez}, {Romani},
  {Roth}, {Ryde}, {Sadrozinski}, {Sanchez}, {Sander}, {Saz Parkinson},
  {Scargle}, {Sgr{\`o}}, {Siskind}, {Smith}, {Smith}, {Spandre}, {Spinelli},
  {Stappers}, {Strickman}, {Suson}, {Tajima}, {Takahashi}, {Tanaka}, {Thayer},
  {Thayer}, {Theureau}, {Thompson}, {Thorsett}, {Tibaldo}, {Torres}, {Tosti},
  {Tramacere}, {Uchiyama}, {Usher}, {Van Etten}, {Vasileiou}, {Vilchez},
  {Vitale}, {Waite}, {Wallace}, {Wang}, {Watters}, {Weltevrede}, {Winer},
  {Wood}, {Ylinen}, \& {Ziegler}}]{abdo+10}
{Abdo}, A.~A., {Ackermann}, M., {Ajello}, M., {et~al.} 2010, \apj, 708, 1254

\bibitem[{{Backer} {et~al.}(2000){Backer}, {Wong}, \& {Valanju}}]{backer+00}
{Backer}, D.~C., {Wong}, T., \& {Valanju}, J. 2000, \apj, 543, 740

\bibitem[{{Brisken} {et~al.}(2010){Brisken}, {Macquart}, {Gao}, {Rickett},
  {Coles}, {Deller}, {Tingay}, \& {West}}]{brisken+10}
{Brisken}, W.~F., {Macquart}, J.-P., {Gao}, J.~J., {et~al.} 2010, \apj, 708,
  232

\bibitem[{{Chevalier}(1977)}]{chevalier+77}
{Chevalier}, R.~A. 1977, \araa, 15, 175

\bibitem[{{Cordes} {et~al.}(2004){Cordes}, {Bhat}, {Hankins}, {McLaughlin}, \&
  {Kern}}]{cordes+04}
{Cordes}, J.~M., {Bhat}, N.~D.~R., {Hankins}, T.~H., {McLaughlin}, M.~A., \&
  {Kern}, J. 2004, \apj, 612, 375

\bibitem[{{Cordes} \& {Rickett}(1998)}]{cordes+98}
{Cordes}, J.~M., \& {Rickett}, B.~J. 1998, \apj, 507, 846

\bibitem[{{Driessen} {et~al.}(2019){Driessen}, {Janssen}, {Bassa}, {Stappers},
  \& {Stinebring}}]{driessen+19}
{Driessen}, L.~N., {Janssen}, G.~H., {Bassa}, C.~G., {Stappers}, B.~W., \&
  {Stinebring}, D.~R. 2019, \mnras, 483, 1224

\bibitem[{{Eilek} \& {Hankins}(2016)}]{eilek+16}
{Eilek}, J.~A., \& {Hankins}, T.~H. 2016, Journal of Plasma Physics, 82,
  635820302

\bibitem[{{Enoto} {et~al.}(2021){Enoto}, {Terasawa}, {Kisaka}, {Hu}, {Guillot},
  {Lewandowska}, {Malacaria}, {Ray}, {Ho}, {Harding}, {Okajima}, {Arzoumanian},
  {Gendreau}, {Wadiasingh}, {Markwardt}, {Soong}, {Kenyon}, {Bogdanov},
  {Majid}, {G{\"u}ver}, {Jaisawal}, {Foster}, {Murata}, {Takeuchi}, {Takefuji},
  {Sekido}, {Yonekura}, {Misawa}, {Tsuchiya}, {Aoki}, {Tokumaru}, {Honma},
  {Kameya}, {Oyama}, {Asano}, {Shibata}, \& {Tanaka}}]{enoto+21}
{Enoto}, T., {Terasawa}, T., {Kisaka}, S., {et~al.} 2021, Science, 372, 187

\bibitem[{{Gelfand} {et~al.}(2009){Gelfand}, {Slane}, \& {Zhang}}]{gelfand+09}
{Gelfand}, J.~D., {Slane}, P.~O., \& {Zhang}, W. 2009, \apj, 703, 2051

\bibitem[{{Gupta} {et~al.}(1999){Gupta}, {Bhat}, \& {Rao}}]{gupta+99}
{Gupta}, Y., {Bhat}, N.~D.~R., \& {Rao}, A.~P. 1999, \apj, 520, 173

\bibitem[{{Gwinn} {et~al.}(1998){Gwinn}, {Britton}, {Reynolds}, {Jauncey},
  {King}, {McCulloch}, {Lovell}, \& {Preston}}]{gwinn+98}
{Gwinn}, C.~R., {Britton}, M.~C., {Reynolds}, J.~E., {et~al.} 1998, \apj, 505,
  928

\bibitem[{{Hankins} \& {Eilek}(2007)}]{hankins+07}
{Hankins}, T.~H., \& {Eilek}, J.~A. 2007, \apj, 670, 693

\bibitem[{{Hankins} {et~al.}(2016){Hankins}, {Eilek}, \& {Jones}}]{hankins+16}
{Hankins}, T.~H., {Eilek}, J.~A., \& {Jones}, G. 2016, \apj, 833, 47

\bibitem[{{Isaacman} \& {Rankin}(1977)}]{isaacman+77}
{Isaacman}, R., \& {Rankin}, J.~M. 1977, \apj, 214, 214

\bibitem[{{Istomin}(2004)}]{istomin04}
{Istomin}, Y.~N. 2004, in IAU Symposium, Vol. 218, Young Neutron Stars and
  Their Environments, ed. F.~{Camilo} \& B.~M. {Gaensler}, 369

\bibitem[{{Jenet} \& {Anderson}(1998)}]{jenet98}
{Jenet}, F.~A., \& {Anderson}, S.~B. 1998, \pasp, 110, 1467

\bibitem[{{Kaplan} {et~al.}(2008){Kaplan}, {Chatterjee}, {Gaensler}, \&
  {Anderson}}]{kaplan+08}
{Kaplan}, D.~L., {Chatterjee}, S., {Gaensler}, B.~M., \& {Anderson}, J. 2008,
  \apj, 677, 1201

\bibitem[{{Karuppusamy} {et~al.}(2010){Karuppusamy}, {Stappers}, \& {van
  Straten}}]{karuppusamy+10}
{Karuppusamy}, R., {Stappers}, B.~W., \& {van Straten}, W. 2010, \aap, 515, A36

\bibitem[{{Lawrence} {et~al.}(1995){Lawrence}, {MacAlpine}, {Uomoto},
  {Woodgate}, {Brown}, {Oliversen}, {Lowenthal}, \& {Liu}}]{lawrence+95}
{Lawrence}, S.~S., {MacAlpine}, G.~M., {Uomoto}, A., {et~al.} 1995, \aj, 109,
  2635

\bibitem[{{Lyne} {et~al.}(1993){Lyne}, {Pritchard}, \&
  {Graham-Smith}}]{lyne+93}
{Lyne}, A.~G., {Pritchard}, R.~S., \& {Graham-Smith}, F. 1993, \mnras, 265,
  1003

\bibitem[{{Lyne} {et~al.}(2001){Lyne}, {Pritchard}, \&
  {Graham-Smith}}]{lyne+01}
---. 2001, \mnras, 321, 67

\bibitem[{{Lyne} \& {Thorne}(1975)}]{lyne+75}
{Lyne}, A.~G., \& {Thorne}, D.~J. 1975, \mnras, 172, 97

\bibitem[{{Main} {et~al.}(2017){Main}, {van Kerkwijk}, {Pen}, {Mahajan}, \&
  {Vanderlinde}}]{main+17}
{Main}, R., {van Kerkwijk}, M., {Pen}, U.-L., {Mahajan}, N., \& {Vanderlinde},
  K. 2017, \apjl, 840, L15

\bibitem[{{Masui} {et~al.}(2015){Masui}, {Lin}, {Sievers}, {Anderson}, {Chang},
  {Chen}, {Ganguly}, {Jarvis}, {Kuo}, {Li}, {Liao}, {McLaughlin}, {Pen},
  {Peterson}, {Roman}, {Timbie}, {Voytek}, \& {Yadav}}]{masui+15}
{Masui}, K., {Lin}, H.-H., {Sievers}, J., {et~al.} 2015, \nat, 528, 523

\bibitem[{{McKee} {et~al.}(2018){McKee}, {Lyne}, {Stappers}, {Bassa}, \&
  {Jordan}}]{mckee+18}
{McKee}, J.~W., {Lyne}, A.~G., {Stappers}, B.~W., {Bassa}, C.~G., \& {Jordan},
  C.~A. 2018, \mnras, 479, 4216

\bibitem[{{Moffett} \& {Hankins}(1996)}]{moffett+96}
{Moffett}, D.~A., \& {Hankins}, T.~H. 1996, \apj, 468, 779

\bibitem[{{Muslimov} \& {Harding}(2004)}]{muslimov+04}
{Muslimov}, A.~G., \& {Harding}, A.~K. 2004, \apj, 606, 1143

\bibitem[{{Osterbrock}(1957)}]{osterbrock57}
{Osterbrock}, D.~E. 1957, \pasp, 69, 227

\bibitem[{{Pen} {et~al.}(2014){Pen}, {Macquart}, {Deller}, \&
  {Brisken}}]{pen+14}
{Pen}, U.-L., {Macquart}, J.-P., {Deller}, A.~T., \& {Brisken}, W. 2014,
  \mnras, 440, L36

\bibitem[{{Petrova}(2004)}]{petrova04}
{Petrova}, S.~A. 2004, \aap, 424, 227

\bibitem[{{Petrova}(2009)}]{petrova09}
---. 2009, \mnras, 395, 1723

\bibitem[{{Philippov} {et~al.}(2019){Philippov}, {Uzdensky}, {Spitkovsky}, \&
  {Cerutti}}]{philippov+19}
{Philippov}, A., {Uzdensky}, D.~A., {Spitkovsky}, A., \& {Cerutti}, B. 2019,
  \apjl, 876, L6

\bibitem[{{Popov} {et~al.}(2017){Popov}, {Rudnitskii}, \&
  {Soglasnov}}]{popov+17}
{Popov}, M.~V., {Rudnitskii}, A.~G., \& {Soglasnov}, V.~A. 2017, Astronomy
  Reports, 61, 178

\bibitem[{{Porth} {et~al.}(2014){Porth}, {Komissarov}, \& {Keppens}}]{porth+14}
{Porth}, O., {Komissarov}, S.~S., \& {Keppens}, R. 2014, \mnras, 443, 547

\bibitem[{{Qiao} {et~al.}(2004){Qiao}, {Lee}, {Wang}, {Xu}, \& {Han}}]{qiao+04}
{Qiao}, G.~J., {Lee}, K.~J., {Wang}, H.~G., {Xu}, R.~X., \& {Han}, J.~L. 2004,
  \apjl, 606, L49

\bibitem[{{Rankin} \& {Counselman}(1973)}]{rankin+73}
{Rankin}, J.~M., \& {Counselman}, III, C.~C. 1973, \apj, 181, 875

\bibitem[{{Ravi} \& {Deshpande}(2018)}]{ravi+18}
{Ravi}, K., \& {Deshpande}, A.~A. 2018, \apj, 859, 22

\bibitem[{{Romani} \& {Yadigaroglu}(1995)}]{romani+95}
{Romani}, R.~W., \& {Yadigaroglu}, I.-A. 1995, \apj, 438, 314

\bibitem[{{Rudnistkii} {et~al.}(2017){Rudnistkii}, {Popov}, \&
  {Soglasnov}}]{rudnistkii+17}
{Rudnistkii}, A.~G., {Popov}, M.~V., \& {Soglasnov}, V.~A. 2017, Astronomy
  Reports, 61, 393

\bibitem[{{Rudnitskii} {et~al.}(2016){Rudnitskii}, {Karuppusamy}, {Popov}, \&
  {Soglasnov}}]{rudnitskii+16}
{Rudnitskii}, A.~G., {Karuppusamy}, R., {Popov}, M.~V., \& {Soglasnov}, V.~A.
  2016, Astronomy Reports, 60, 211

\bibitem[{{Shearer} {et~al.}(2003){Shearer}, {Stappers}, {O'Connor}, {Golden},
  {Strom}, {Redfern}, \& {Ryan}}]{shearer+03}
{Shearer}, A., {Stappers}, B., {O'Connor}, P., {et~al.} 2003, Science, 301, 493

\bibitem[{{Shklovsky}(1957)}]{shklovsky57}
{Shklovsky}, I.~S. 1957, in IAU Symposium, Vol.~4, Radio astronomy, ed. H.~C.
  {van de Hulst}, 201

\bibitem[{{Slane}(2017)}]{slane17}
{Slane}, P. 2017, {Pulsar Wind Nebulae}, ed. A.~W. {Alsabti} \& P.~{Murdin},
  2159

\bibitem[{{Smirnova} {et~al.}(1996){Smirnova}, {Shishov}, \&
  {Malofeev}}]{smirnova+96}
{Smirnova}, T.~V., {Shishov}, V.~I., \& {Malofeev}, V.~M. 1996, \apj, 462, 289

\bibitem[{{Stinebring} {et~al.}(2001){Stinebring}, {McLaughlin}, {Cordes},
  {Becker}, {Goodman}, {Kramer}, {Sheckard}, \& {Smith}}]{stinebring+01}
{Stinebring}, D.~R., {McLaughlin}, M.~A., {Cordes}, J.~M., {et~al.} 2001,
  \apjl, 549, L97

\bibitem[{{Strader} {et~al.}(2013){Strader}, {Johnson}, {Mazin}, {Spiro
  Jaeger}, {Gwinn}, {Meeker}, {Szypryt}, {van Eyken}, {Marsden}, {O'Brien},
  {Walter}, {Ulbricht}, {Stoughton}, \& {Bumble}}]{strader+13}
{Strader}, M.~J., {Johnson}, M.~D., {Mazin}, B.~A., {et~al.} 2013, \apjl, 779,
  L12

\bibitem[{{Vandenberg}(1976)}]{vandenberg+76b}
{Vandenberg}, N.~R. 1976, \apj, 209, 578

\bibitem[{{Vandenberg} {et~al.}(1976){Vandenberg}, {Clark}, {Erickson},
  {Resch}, \& {Broderick}}]{vandenberg+76}
{Vandenberg}, N.~R., {Clark}, T.~A., {Erickson}, W.~C., {Resch}, G.~M., \&
  {Broderick}, J.~J. 1976, \apj, 207, 937

\bibitem[{{Wolszczan} \& {Cordes}(1987)}]{wolszczan+87}
{Wolszczan}, A., \& {Cordes}, J.~M. 1987, \apjl, 320, L35

\end{thebibliography}

\end{document}